\documentclass[fleqn,usenatbib]{mnras}

\usepackage[T1]{fontenc}

\DeclareRobustCommand{\VAN}[3]{#2}
\let\VANthebibliography\thebibliography
\def\thebibliography{\DeclareRobustCommand{\VAN}[3]{##3}\VANthebibliography}

\usepackage{graphicx}	
\usepackage{amsmath}	
\usepackage{amssymb}	
\usepackage{hyperref}
\usepackage[dvipsnames]{xcolor}
\hypersetup{
    colorlinks = true,
    linkcolor = NavyBlue,
    urlcolor = NavyBlue
    }

\usepackage{newtxtext,newtxmath}

\title[Constraints from GW spin-orbit misalignment]{Constraints on compact binary merger evolution from spin-orbit misalignment in gravitational-wave observations}

\author[B. P. Gompertz et al.]{
B. P. Gompertz,$^{1}$\thanks{E-mail: b.gompertz@bham.ac.uk (BPG)}
M. Nicholl,$^{1}$
P. Schmidt,$^{1}$
G. Pratten,$^{1}$
and A. Vecchio$^{1}$
\\
$^{1}$School of Physics and Astronomy \& Institute for Gravitational Wave Astronomy, University of Birmingham, Birmingham, B15 2TT, UK\\
}

\date{Accepted XXX. Received YYY; in original form ZZZ}

\pubyear{2021}

\begin{document}
\label{firstpage}
\pagerange{\pageref{firstpage}--\pageref{lastpage}}
\maketitle

\begin{abstract}
The identification of the first confirmed neutron star - black hole (NS-BH) binary mergers by the LIGO, Virgo and KAGRA collaboration provides the opportunity to investigate the properties of the early sample of confirmed and candidate events. Here, we focus primarily on the tilt angle of the black hole's spin relative to the orbital angular momentum vector of the binary, and the implications for the physical processes that determine this tilt. The posterior tilt distributions of GW200115 and the candidate events GW190426\_152155 and GW190917\_114630 peak at significantly anti-aligned orientations (though display wide distributions). Producing these tilts through isolated binary evolution would require stronger natal kicks than are typically considered (and preferentially-polar kicks would be ruled out), and/or an additional source of tilt such as stable mass transfer. The early sample of NS-BH events are less massive than expected for classical formation channels, and may provide evidence for efficient mass transfer that results in the merger of more massive NS-BH binaries before their evolution to the compact phase is complete. We predict that future gravitational-wave detections of NS-BH events will continue to display total binary masses of $\approx 7$\,M$_{\odot}$ and mass ratios of $q \sim 3$ if this interpretation is correct. Conversely, the high mass of the candidate GW191219\_163120 suggests a dynamical capture origin. Large tilts in a significant fraction of merging NS-BH systems would weaken the prospects for electromagnetic detection. However, EM observations, including non-detections, can significantly tighten the constraints on spin and mass ratio.
\end{abstract}




\section{Introduction}
The era of gravitational-wave (GW) astronomy has driven a revolution in our understanding of the physics of compact objects. Since the first discovery of a merging binary black hole (BBH) system \citep{Abbott16}, the Advanced LIGO \citep{LIGO15} and Virgo \citep{Acernese15} interferometers have provided a remarkable wealth of information on the denizens of the `stellar graveyard'. Particularly notable milestones include the identification of the first intermediate mass black hole \citep[BH;][]{Abbott20b} and the detection of the landmark GW170817, the first signal from a merging binary neutron star (NS) system \citep{Abbott17}. The latter event also yielded the first ever electromagnetic (EM) counterpart to a GW event; the gamma-ray burst (GRB) 170817A \citep{Abbot17_GRB,Goldstein17,Hallinan17,Margutti17,Savchenko17,Troja17,DAvanzo18,Lyman18,Margutti18,Mooley18,Troja18,Lamb19}, and the kilonova (KN) AT2017gfo \citep{Andreoni2017,Arcavi2017,Chornock17,Coulter17,Cowperthwaite17,Drout17,Evans17,Kasliwal2017,Lipunov2017,McCully2017,Nicholl17,Pian17,Shappee2017,Smartt17,Soares-Santos17,Tanvir17,Utsumi2017,Valenti2017,Villar17}, heralding the advent of GW-EM multi-messenger astronomy \citep{Abbott17b}.

More recently, the LIGO-Virgo-KAGRA \citep{KAGRA19} collaboration (LVK) announced the detection of the merger of two NS-BH binary mergers \citep{Abbott21}, completing the set of compact binary merger constituents. The two binaries were measured to have component masses of $8.9^{+1.2}_{-1.5}$\,M$_{\odot}$ and $1.9^{+0.3}_{-0.2}$\,M$_{\odot}$ (GW200105), and $5.7^{+1.8}_{-2.1}$\,M$_{\odot}$ and $1.5^{+0.7}_{-0.3}$\,M$_{\odot}$ (GW200115) assuming agnostic priors for the spin of the secondary. Interestingly, the spin projection of the primary of GW200115 onto the orbital angular momentum was negative with 88 per cent probability, potentially indicating a significant spin-orbit tilt angle. The primary spin magnitude of GW200105 was less than $0.23$, but its direction was unconstrained. In addition to the newly-confirmed events, known NS-BH candidates also include GW190426\_152155 and GW190814 \citep{Abbott21b}, GW190917\_114630 \citep{GWTC-2.1}, and GW191219\_163120 \citep{GWTC-3}. The $2.59^{+0.09}_{-0.08}$\,M$_{\odot}$ secondary of GW190814 \citep{Abbott20} is much more massive than even the heaviest known NSs \citep{Demorest10,Antoniadis13,Cromartie20} and is omitted here as a likely BBH event.

The NS-BH variant of compact object mergers garners particular attention because it enables observers to explore the maximum stable mass of an NS, and the lower `mass gap' between $3$--$5$\,M$_{\odot}$ \citep[e.g.][]{Ozel10,Farr11}. If the inspiraling NS is disrupted before plunging into the BH, EM observations may also be possible \citep[e.g.][]{Kawaguchi16,Foucart18,Fernandez20,Kruger20}. Combined with GW measurements of the mass, EM observations of compact binary mergers can provide multi-messenger constraints on the NS equation-of-state (EoS) \citep{Margalit2017,Margalit19,Coughlin2019,Dietrich2020,Nicholl21,Breschi2021}.

The ability to study non-photonic events, previously inaccessible to traditional EM astronomy, has also led to tighter constraints on properties such as merger rates, mass distributions, and binary configurations \citep{GWTC2pop}. In turn, these observations can be used to elucidate the properties of the population of compact binaries prior to the merger. In particular, the mechanism by which the binary was formed may imprint itself on the tilt angle $\theta_{\rm BH}$; the inclination between the BH spin and the binary orbital plane. This possibility has been previously explored in e.g. \citet{Farr17,Stevenson17,Talbot:2017yur,Vitale17,Roulet21}.

In cases where the binary formed via dynamical capture, whether in dense stellar clusters \citep[e.g.][]{Hoang20,Rastello20,Ye20} or AGN discs \citep[e.g.][]{McKernan20,McKernan20b}, we can expect an isotropic distribution of tilt angles because the spins of the two constituents are independent of one another and the orbital plane is set by the direction of approach during the capture \citep[e.g.][]{Rodriguez16}. Where the system was formed via isolated binary evolution \citep[e.g.][]{Smarr1976,Srinivasan1989,Belczynski2002}, the misalignment is expected to be small because viscous forces will align both the spins and the orbit along the dominant angular momentum plane of the original gas cloud.

However, the identification of significant tilt angles is not necessarily a smoking gun for dynamical capture because several mechanisms still exist by which the spin of one or both of the constituents of a binary formed in isolation may become misaligned with the orbital plane. Possible mechanisms include natal kicks that imbue a significant velocity on the stellar remnant during core collapse \citep[e.g.][]{Rodriguez16}, or stable mass transfer that may tilt the donor star's spin vector into the orbital plane \citep[e.g.][]{Stegmann21}. In addition to this, the efficiency with which common envelope (CE) processes and tidal forces can realign the spin and orbit vectors is uncertain.

\citet{Broekgaarden21b} demonstrated that the chirp masses, component masses and mass ratios of GW200105 and GW200115 are consistent with an isolated binary evolution channel in their population synthesis modelling. Combined with the low expected rates of dynamical capture-driven NS-BH mergers \citep[e.g.][]{ArcaSedda20,Hoang20,Ye20}, we consider the isolated binary formation channel as our preferred pathway. In this paper, we investigate the physical processes that may produce significant spin-orbit anti-alignment, and compare the tilt angles of confirmed or candidate NS-BH mergers to the population of BBH mergers presented in \citep{GWTC2pop}. In Section~\ref{sec:GW_observations} we describe the data sample and assess the GW posteriors for the BH tilt angle. We then draw constraints on compact binary evolution in Section~\ref{sec:formation_channels} and their implications for future observations of compact binaries in Section~\ref{sec:implications}. Finally, we discuss our conclusions in Section~\ref{sec:conclusions}.

\section{Gravitational-wave observations}
\label{sec:GW_observations}
To determine whether the BH spin tilt angles $\theta_{\rm BH}$ observed in the NS-BH mergers are consistent with the observed astrophysical BH population, we compare the BH tilt angle measurements of the NS-BH events to the full GW-detected BH sample, which consists exclusively of BBH mergers. We construct posterior predictive distributions (PPD) for the spin tilt using the public posteriors from the default spin model \citep{GWTC2pop,GWTC2popDR,GWOSC19}, treating the spin of each component BH as being independently drawn from the same underlying distribution. The default spin tilt distribution is a mixture model consisting of an isotropic component and a second component in which the spins are preferentially (anti-) aligned with the orbital angular momentum \citep{Talbot:2017yur}. We note that only the BHs listed in Table 1 of \citet{GWTC2pop} are included in this population analysis.

\begin{figure}
\includegraphics[width=\columnwidth]{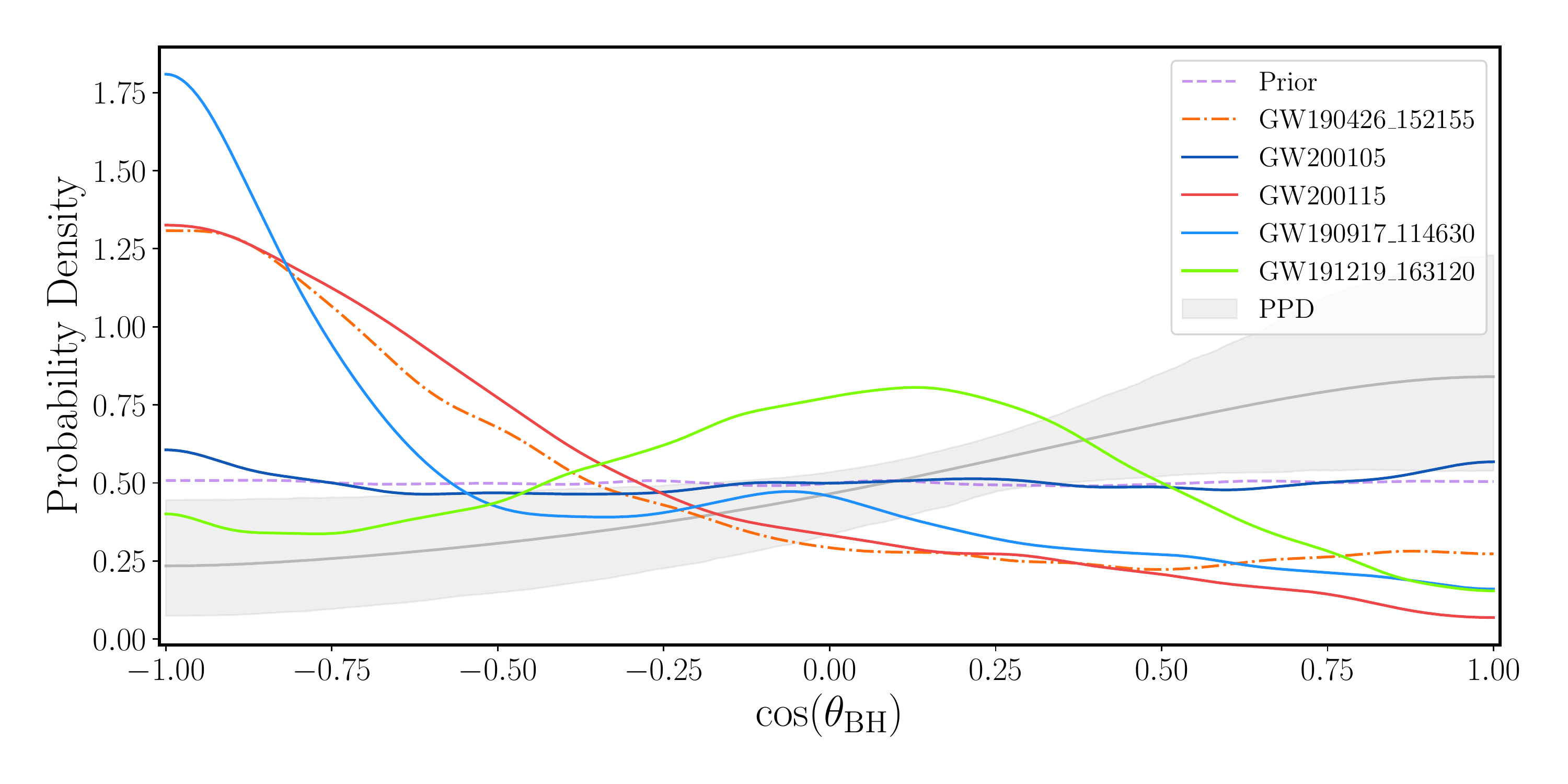}
\caption{Posterior predictive distribution (PPD) of the BH tilt angles assuming the default spin model of \citet{GWTC2pop}. The BH spins of GW200115 (red), GW190426\_152155 (orange dashed) and GW190917\_114630 (light blue) clearly stand out from the BH population (grey), where the solid grey line is the mean of the PPD and the shaded region demarcates the 5th and 95th quantile.}
\label{fig:BBH_PPD}
\end{figure}

Figure~\ref{fig:BBH_PPD} shows the one-dimensional posterior probability density (PDF) of the cosine of the BH tilt angle for the population of BBH mergers (grey) alongside the tilt PDFs of the BHs in the individual NS-BH events. We take our posterior sample sets from the most recent LVK data release available (Table~\ref{tab:data}). While the BH spin tilts of GW191219\_163120 and GW200105 are indistinguishable from the prior, the BH spin tilts of GW190426\_152155, GW190917\_114630 and GW200115 show posterior support that is distinct from the prior and the observed BH population. Since the spin measurements are dominated by the primary object and the tilt angle is expected to be largely unchanged after the first supernova (SN) \citep[e.g.][]{Gerosa18}, such a difference may reflect discrepancies in either the pre-SN evolution or the magnitude of the natal kicks experienced by the two populations. In particular, BHs at the higher end of the BBH mass distribution likely come from distinct progenitor populations, and form via direct collapse with very little mass ejection and highly suppressed natal kicks.

For a quantitative measure of the statistical difference between the mean BH tilt PPD and the one-dimensional PDF for the five NS-BH candidates, we employ the Jensen-Shannon divergence \citep[$D_{\rm JS}$;][see Table~\ref{tab:data}]{Lin91divergencemeasures}. We find a departure between the measured spin tilt posteriors of GW200115, GW190917\_114630 and GW190426\_152155 and the median of the observed astrophysical BH population from the BBH events. The median values of the spin tilt posteriors for these three events are $\approx -0.55$, compared to $+0.24$ for the BBH population. This suggests support for the hypothesis that the BHs of at least GW200115, GW190917\_114630 and GW190426\_152155 stand out from the observed population. The BH tilt posteriors for GW200105 and GW191219\_163120, on the other hand, are uninformative relative to the spin tilt prior and fully consistent with the population PPD. The cumulative distribution functions (CDFs) of GW200115, GW190917\_114630 and GW190426\_152155 show $79.9$, $72.6$ and $74.2$ per cent of the total probability in favour of tilts greater than 90 degrees ($\cos(\theta) \leq 0$), respectively.

It is important to note here that all of the statements we make about the posterior distributions are sensitive to the choice of priors \citep[see e.g.][]{Vitale17b,Mandel20,Zevin20}. These are uniform in spin magnitude and isotropic in spin orientation \citep{Abbott21b}. Different assumptions about the prior may produce different posterior distributions, though quantifying this effect is beyond the scope of this paper.

\begin{table*}
    \centering
    \begin{tabular}{cccccccc}
        \hline\hline
        Event & Tilt & $D_{\rm JS}$ & m$_1$ & m$_2$ & m$_{\rm tot}$ & Release & Sample set \\
         & $ \cos{(\theta)}$ & & M$_{\odot}$ & M$_{\odot}$ & M$_{\odot}$ & \\
         \hline
        GW190426\_152155 & $-0.53^{+1.36}_{-0.43}$ & $0.17$ & $5.7^{+3.9}_{-1.5}$ & $1.5^{+0.8}_{-0.5}$ & $7.2^{+3.5}_{-1.5}$ & GWTC-2 & {\sc PrecessingSpinIMRHM} \\
        GW190917\_114630 & $-0.54^{+1.27}_{-0.44}$ & $0.17$ & $9.3^{+3.4}_{-4.4}$ & $2.1^{+1.5}_{-0.5}$ & $11.4^{+3.0}_{-2.9}$ & GWTC-2.1 & {\sc PrecessingSpinIMRHM} \\
        GW191219\_163120 & $0.00^{+0.74}_{-0.86}$ & $0.07$ & $31.1^{+2.2}_{-2.8}$ & $1.17^{+0.07}_{-0.06}$ & $32.3^{+2.2}_{-2.7}$ & GWTC-3 & {\sc C01:Mixed} \\
        GW200105 & $0.01^{+0.90}_{-0.93}$ & $0.03$ & $9.0^{+1.7}_{-1.7}$ & $1.91^{+0.33}_{-0.24}$ & $11.0^{+1.5}_{-1.4}$ & GWTC-3 & {\sc C01:Mixed} \\
        GW200115 & $-0.56^{+1.15}_{-0.40}$ & $0.22$ & $5.9^{+2.0}_{-2.5}$ & $1.44^{+0.85}_{-0.29}$ & $7.4^{+1.8}_{-1.7}$ & GWTC-3 & {\sc C01:IMRPhenomXPHM:HighSpin} \\
        
        \hline\hline
    \end{tabular}
    \caption{GW events taken to be NS-BH merger candidates in this analysis. Data presented are median values with 90 per cent credible intervals.}
    \label{tab:data}
\end{table*}

\section{Constraints on compact binary evolution}
\label{sec:formation_channels}

The dominant channel for producing compact binary mergers is expected to be isolated binary evolution, wherein the two stars are born, evolve, and merge without outside interaction.  This channel predicts merger rates of $\sim {\rm few}\times1-100$\,Gpc$^{-3}$\,yr$^{-1}$, consistent with the current population of GW-detected NS-BH mergers \citep{Dominik2013,Mapelli2018,Wysocki2018,Broekgaarden21,Broekgaarden21b}. Rates lower by 1-2 orders of magnitude are predicted from formation in globular clusters \citep{Clausen2013} or from isolated triples \citep{Fragione2019}. However, given current model uncertainties, significant contributions to the merger rate could also arise from dynamical interactions in young stellar clusters (\citealt{Rastello20,Santoliquido2020}; but see \citealt{Fragione2020,Hoang20}) and the disks of active galactic nuclei \citep{McKernan2020}.

Broadly, the `classic' isolated binary channel proceeds as follows \citep[e.g.][]{Smarr1976,Srinivasan1989,Belczynski2002}: the binary is born with the component spins roughly aligned with the orbit. The more massive (primary) component evolves to fill its Roche lobe, and initiates stable mass transfer onto the secondary. Eventually, the primary star ends its life as a core-collapse SN and forms the first compact object, which is usually a black hole. The SN imparts a natal kick on the primary, which can unbind over 95 per cent of wide binaries \citep[e.g.][]{Renzo19}. This natal kick is thought to be the dominant influence on the eventual tilt of the primary \citep{Gerosa18}, with the tilt largely unchanged from this point onward.

If the binary survives the kick, the secondary star will later evolve to fill its Roche lobe and initiate mass transfer onto the primary. This transfer can become unstable, and result in a common envelope (CE) phase \citep[e.g.][]{Ivanova13}. The CE rapidly decreases the binary separation through viscous drag. If the CE can be ejected before a merger occurs, the subsequent core-collapse SN of the secondary results in a tight compact object binary that will eventually merge as its orbital angular momentum is lost to GW emission.

\subsection{Natal kicks}

In the absence of direct observations of BH velocities, the distribution of their natal kicks is often assumed to be a modified version of the NS natal kick distribution. This is typically taken to be a Maxwellian distribution with a dispersion of $\sigma = 265$\,km/s, based on the observed velocities of pulsars in the Milky Way \citep{Hobbs05}. \citet{Rodriguez16} explored the BBH tilt angles induced by natal kicks with three different prescriptions for converting the observed NS kick distribution into a BH kick distribution:
\begin{enumerate}
    \item ``Fallback kicks'', wherein some fraction $f_{\rm fb}$ of the ejected mass (proportional to the core mass of the progenitor star) will ``fall back'' onto the BH, damping its velocity by $V_{\rm kick}^{\rm BH} = (1 - f_{\rm fb})V_{\rm kick}^{\rm NS}$.
    \item ``Proportional kicks'', where the BH natal kick is modified in proportion with the maximum mass of an NS by $V_{\rm kick}^{\rm BH} = \big(\frac{m_{\rm NS}}{m_{\rm BH}}\big)V_{\rm kick}^{\rm NS}$.
    \item ``Full kicks'', where $V_{\rm kick}^{\rm BH} = V_{\rm kick}^{\rm NS}$.
\end{enumerate}

Each prescription was assessed for an isotropic distribution of kicks, and a `polar' distribution confined to a cone with a half opening angle of $10^{\circ}$ from the star's spin axis. Polar kicks are motivated by the observed relationship between the rotation axis and direction of motion in some pulsars \citep[e.g.][]{Johnston05,Yao21}. The BBH population was created by evolving a stellar population with the Binary Stellar Evolution \citep[BSE;][]{Hurley02} code. This parent population featured a range of metallicities between $0.005$\,Z$_{\odot}$ and $1.5$\,Z$_{\odot}$. Primary masses were drawn from a \citet{Kroupa01} initial mass function ranging between $18$\,M$_{\odot}$ and $150$\,M$_{\odot}$, with a flat mass ratio distribution between 0 and 1. The semi-major axes of the binaries were drawn from a flat distribution in log space between $10$\,R$_{\odot}$ and $10^5$\,R$_{\odot}$. The final sample of BBHs was limited to only those that would merge via gravitational-wave emission within $13.8$\,Gyr. The tilt of the orbit after each natal kick is calculated using the formalism developed in the appendix of \citet{Hurley02}. After being kicked, each binary was evolved \citep{Gerosa16} until it entered the LIGO band (10\,Hz), at which point the tilts are assessed. For further discussion on the effects of natal kicks on binary orbits and how spin-orbit tilt can be calculated, see e.g. \citet{Tauris98,Brandt95,Kalogera96,Kalogera00}.

\begin{figure}
\includegraphics[width=\columnwidth]{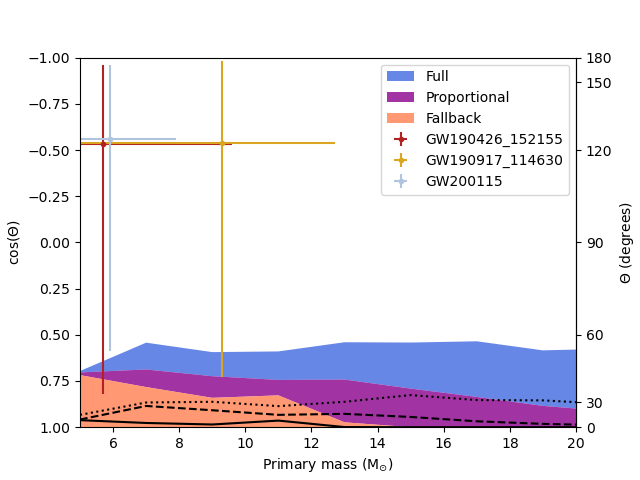}
\includegraphics[width=\columnwidth]{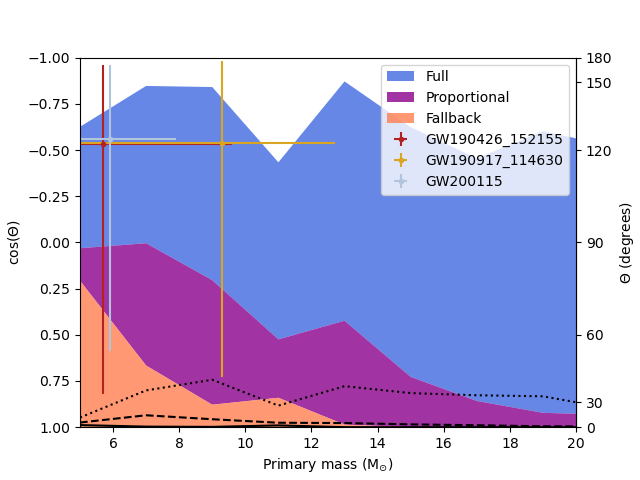}
\caption{The spin-orbit tilt angles induced on a nascent BH as a result of the natal kick it receives during SN as a function of its mass for primary stars considered in \citet{Rodriguez16}. Shaded regions show where the lower 90 per cent of sources lie for three different kick prescriptions, with the median tilts for each indicated by the solid, dashed and dotted lines for the fallback, proportional and full kicks, respectively. Data are grouped into bins of $2$\,M$_{\odot}$.} The upper panel is for polar kicks, and the lower for an isotropic distribution. The 90 per cent confidence intervals for the primary mass and tilt is shown for GW200115, GW190917\_114630 and GW190426\_152155.
\label{fig:tilts}
\end{figure}

Using the data of \citet{Rodriguez16}, we investigate the tilt of the primary star as a function of its mass for each kick prescription (Figure~\ref{fig:tilts}). Data are grouped into bins of $2$\,M$_{\odot}$, which is the cause of the `spiky' curve. Polar kicks produce misalignments of $\cos(\theta) > 0.5$ in almost all cases, including for full kicks. While only the ``isotropic full kicks'' prescription produces significant tilt across the full range of masses, the ``isotropic fallback'' and ``isotropic proportional'' kick prescriptions can produce significant, even anti-aligned, tilts for lower mass stars. If tidal forces can efficiently realign the spins with the orbital plane, the expected tilt angles are greatly reduced \citep{Rodriguez16}. However, tidal processes will preferentially work to align the spin of the less massive object, which is usually the second SN \citep{Gerosa18}, whereas here we are interested in the more massive object (since it is the tilt of the BH that is well measured).

We consider the three events that show significant posterior divergence from both the prior and the BBH tilt distribution (Table~\ref{tab:data}). The 90 per cent confidence interval of the tilt of GW200115 is inconsistent with the 90th centile of tilts for an isolated binary evolution channel that favours polar natal kicks, regardless of which prescription is used. GW190917\_114630 is inconsistent with the proportional and fallback tilt distributions. The 90 per cent confidence interval of the tilt of GW190426\_152155 does marginally overlap with the 90th centile of tilts for the polar kick distributions. Considering the three events together, we conclude that a polar kick distribution for BHs is disfavoured by the GW observations. Although we are comparing NS-BH mergers to BBH mergers here, Figure~\ref{fig:tilts} demonstrates that tilt can be achieved via natal kicks in lower mass BHs, while in higher mass cases these natal kicks are suppressed either by fallback material or conservation of momentum arguments. We expect BBH primaries with masses comparable to the BHs in our NS-BH events to be directly comparable. This may not be the case for higher mass systems, which could arise through different evolutionary pathways.

When compared to the isotropic kick distributions (Figure~\ref{fig:tilts}, lower), the peak probabilities of the posterior tilt distributions of GW200115, GW190917\_114630 and GW190426\_152155 are towards the edges of the $90^{\rm th}$ centile of tilts produced by BH natal kicks, although their confidence intervals extend to more moderate tilts. GW190917\_114630 is inconsistent with the fallback distribution. Having three of the first five known events (with the other two unconstrained) towards the edge of the expected tilt distribution is not impossible, but it may hint at either another mechanism by which tilt can be induced, or point towards higher-than-expected BH natal kick velocities. The latter conclusion was also suggested to explain the tilt of the low-mass BBH GW151226 \citep{OShaugnessy17}, reconcile the relative rates of BBH and binary NS mergers detected in GW \citep{Chruslinska18}, align the predicted and observed spin distributions of the BBH population \citep{Callister20} and in independent analyses of GW200105 and GW200115 \citep{Fragione21}. Some estimates of BH velocities from low mass X-ray binaries imply velocities of 80\,km\,s$^{-1}$ or less \citep{Mandel_BH_kicks}, though other studies prefer higher kick velocities of $> 100$\,km\,s$^{-1}$ \citep{Repetto17}. Alternatively, \citet{Zhu21} suggest that large tilt angles could be achieved in GW200115 with an extremely high velocity ($600$\,km/s) natal kick to the NS during the second SN in the binary if it were directed within $30^{\circ}$ of the pre-SN orbital plane.

\subsection{Additional tilt from stable mass transfer}\label{sec:mass_transfer}

\citet{Stegmann21} explore the effects of stable mass transfer prior to the first SN on the tilt angle of BBH systems. Mass transfer is shown to slowly tilt the donor (primary) star from an (assumed) initially aligned configuration ($\cos(\theta) = 0.9$ - $1.0$) until the spin is significantly misaligned ($\cos(\theta) \sim 0$). End-to-end simulations accounting for mass transfer, natal kicks, tidal forces, and the CE phase result in a population of BBH mergers with a tilt angle distribution that peaks at an aligned configuration, but remains relatively flat between $\cos(\theta) = 1$ and $\cos(\theta) = 0$. If tides are weak, this distribution may even become bimodal around $\cos(\theta) = 1$ and $\cos(\theta) = 0$. Anti-aligned spins ($\cos(\theta) < 0$) remain comparatively rare; an order of magnitude or more less common.

The median tilts of our sample with significant divergence between posterior and prior (Table~\ref{tab:data}) lie well away from the peaks of the tilt angle probability distributions discussed in \citet{Stegmann21}. However, the 90 per cent confidence intervals are broad, and consistent with $\cos(\theta) = 0$, where the probability rapidly rises. The inclusion of tilt via stable mass transfer therefore serves to ease the tension between the expectations for BH kicks and the peak posterior measurements. However, the peaks of the distributions still favour tilts that are greater than would be expected from a combination of canonical BH kick velocities and tilting via stable mass transfer cf. \citet{Stegmann21}.

\subsection{Observational implications for binary evolution physics}\label{sec:obs_implications}

Independent of its potential to induce tilt, stable mass transfer prior to the first SN has important effects on the mass distribution and merger rate of NS-BH binaries \citep{Kruckow18,Broekgaarden21,Shao21}. In particular, when accretion of the mass lost by the primary onto the secondary is efficient, binaries following the `classic' channel (see Section~\ref{sec:formation_channels}) tend to merge before they can become compact binaries \citep{Broekgaarden21}. This leads to a reduction of the overall NS-BH merger rate, and a rise in the relative rates of non-traditional channels such as those that avoid CE (or alternatively experience it twice), or those that proceed via chemically homogeneous evolution \citep[e.g.][]{Marchant17}.

We emphasize that the predictions for a higher efficiency stable mass transfer pathway are in good agreement with the observations of the first two confirmed NS-BH mergers, and the candidates GW190426\_151255 and GW190917\_114630. In particular, Model C in \citet{Broekgaarden21}, which fixes the fraction of mass lost by the donor that is accreted by the secondary to $\beta = 0.5$, predicts that the GW-detected distribution of NS-BH mergers will peak at a total binary mass of $\approx 7$\,M$_{\odot}$, with a preference for mass ratios of $q \approx 3$. This compares favourably to our sample (Table~\ref{tab:data}), particularly GW190426\_151255 and GW200115. The exception is GW191219\_163120, which has a $31.1$\,M$_{\odot}$ BH that is not well explained by any of the progenitor pathway models. Furthermore, the event-based NS-BH merger rate derived in \citet{Abbott21} from GW200105 and GW200115 is fully consistent with the expected merger rate of Model C in \citet{Broekgaarden21}, although both exhibit broad distributions.

The early NS-BH merger observations may therefore suggest that the `classic CE' formation channel where binaries undergo one CE phase may be less common than expected in NS-BH events. If orbital decay is indeed being facilitated in many cases by efficient stable mass transfer or two common envelope phases, then future NS-BH merger events (e.g. during the LVK O4 run) will continue to show total binary masses distributed around $7$\,M$_{\odot}$. This is in contrast to the mass distribution in the fiducial model of \citet{Broekgaarden21}, which peaks at around $15$\,M$_{\odot}$ (in part because more massive systems should be easier to detect in GW). A preference for lower mass ($5$ -- $15$\,M$_{\odot}$) BHs (and higher mass NSs), compatible with the observed NS-BH sample, was also shown in \citet{Giacobbo2018}.

Conversely, some dynamical capture channels suggest NS-BH mergers that are significantly more massive than expected for the isolated binaries \citep[e.g.][]{Rastello20}. This pathway may provide a natural explanation for GW191219\_163120. \citet{Rastello20} find that roughly 25 per cent of NS-BH mergers resulting from dynamical capture will have a total binary mass of $> 30$\,M$_{\odot}$. They estimate the overall rate of mergers formed in this way to be $\sim 28$\,Gpc$^{-3}$\,yr$^{-1}$, a factor of $\sim 2$ less than the isolated evolution merger rate of $\sim 49$\,Gpc$^{-3}$\,yr$^{-1}$ \citep{Santoliquido2020}. Therefore, our relative observed rates of four NS-BH mergers consistent with isolated binary evolution to one very massive ($m_{\rm tot} > 30$\,M$_{\odot}$) NS-BH merger potentially driven by dynamical capture is in good agreement with the rates derived by \citet{Rastello20} and \citet{Santoliquido2020}.

Predictions from population synthesis modelling for the spin magnitude of the BH at the point of merger are uncertain. A common prediction for BBH evolution is that a BH will have essentially zero spin at the point of merger, having lost most of its angular momentum to stellar winds, mass transfer or angular momentum transport at the point of collapse \citep[e.g.][]{Qin18,Fuller19}. It should be noted that despite supporting anti-aligned tilts, the broad posterior spin distributions of GW190426\_152155, GW190917\_114630 and GW200115 are consistent with this picture. However, highly-spinning and anti-aligned BBHs are expected in cases with weak stellar core-envelope coupling or through tidal synchronisation \citep{Steinle21}. A BH may also be spun up prior to its collapse if it is born second, resulting in a large spin at the point of merger \citep{Bavera20,Chattopadhyay21}. Ultimately, high mass X-ray binaries \citep{Miller15} and long GRBs \citep[e.g.][]{Fryer19} both provide evidence that some BHs must be rapidly rotating in binary systems at some point in their evolution. The true picture is awaiting observational confirmation, although given the degeneracies in the GW posteriors, associated EM constraints will likely be required to uniquely identify the characteristics of the pre-merger binary.

\section{Implications for multi-messenger events}\label{sec:implications}

NS-BH mergers are a potential source of EM emission if the NS is disrupted during inspiral, such that some of the material remains outside of the BH event horizon. Hot, dense, and neutron-rich material could facilitate rapid neutron capture \citep[r-process;][]{Lattimer74,Eichler89,Freiburghaus99} nucleosynthesis, which would lead to a thermal transient known as a kilonova \citep[KN;][]{Li98,Rosswog05,Metzger10,Barnes13,Metzger17} when the newly-formed unstable heavy elements flow away from the merger site in winds and dynamical ejecta, and decay to stability. If a few tenths of a solar mass accrete onto the BH then the merger may also power a short GRB, if a relativistic jet can be launched successfully \citep{Blandford77,Barbieri19,Gompertz20}.

\begin{figure}
\includegraphics[width=9cm]{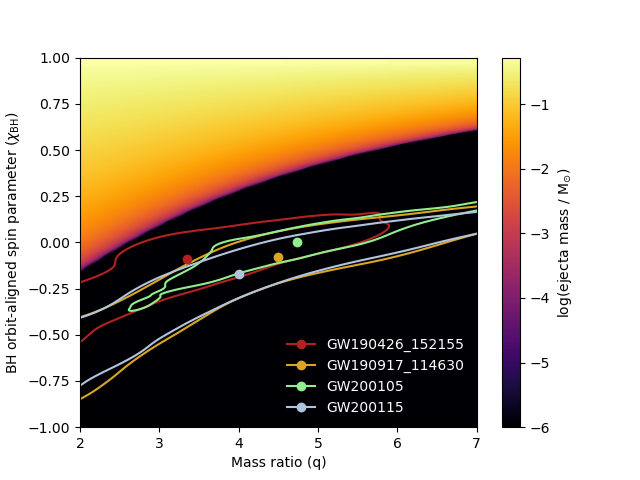}
\caption{The post-merger remnant mass as a function of the orbit-aligned component of the BH spin ($\chi_{\rm BH}$) and the binary mass ratio ($q$) \citep[cf.][]{Foucart18}. An NS with $m_2 = 1.4$\,M$_{\odot}$ and $r = 12$\,km is assumed. Contours show the 90 per cent credible intervals for the NS-BH events and candidates.}
\label{fig:ejecta_mass}
\end{figure}

The dominant properties that predict the remnant mass outside of the BH event horizon are the orbit-aligned component of the spin of the pre-merger BH ($\chi_{\rm BH}$) and the mass ratio of the binary, $q$. The compactness of the NS is also important. Figure~\ref{fig:ejecta_mass} shows the expected remnant mass \citep[cf.][]{Foucart18} as a function of $\chi_{\rm BH}$ and $q$, assuming an NS with $M = 1.4$\,M$_{\odot}$ and a radius of 12\,km. Essentially no mass is expected to remain outside of the event horizon for any of our NS-BH candidates, making the prospect of an EM accompaniment unlikely. We note that follow-up of these four events during O3 did not yield any detected counterparts \citep{Hosseinzadeh2019,Lundquist2019,Gompertz2020,Paterson2020,Anand2021}. While this outcome is in line with the expected absence of an EM transient from modelling \citep[see also][]{Zhu21b}, efforts to constrain one were also hindered by a combination of poor localisations ({1300, 2100, 7700, 900} square degrees, 90 per cent containment) and large distances ({$370^{+320}_{-300}$, $720^{+340}_{-310}$, $280^{+110}_{-110}$, $300^{+150}_{-100}$}\,Mpc) for GW190426\_152155, GW190917\_114630, GW200105 and GW200115 respectively \citep{Abbott21,Abbott21b,GWTC-2.1}.

If NS-BH mergers do show low $\chi_{\rm BH}$, whether due to a low spin magnitude or large tilt, the prospects for NS-BH-driven KNe and GRBs are poor. However, Figure~\ref{fig:ejecta_mass} demonstrates that with a modest increase in the GW-measured spin, EM observations could significantly tighten the constraints on $\chi_{\rm BH}$ and $q$ by independently detecting or ruling out a transient. This possibility has important implications for constraining the properties of nuclear matter through multi-messenger observations even where no detection is made. Furthermore, the presence or absence of EM-bright NS-BH mergers provides an immediate constraint on their average spins, and therefore their progenitor pathways, as discussed in Section~\ref{sec:obs_implications}.

\section{Conclusions}\label{sec:conclusions}

We investigate the tilt angles of the recently announced NS-BH merger events GW200105 and GW200115 \citep{Abbott21} and the candidates GW190426\_152155, GW190917\_114630 and GW191219\_163120, and compare them to the tilt angles derived for the known population of BBH events \citep{GWTC2pop}. The posterior tilt distributions of GW200115, GW190917\_114630 and GW190426\_152155 place them in a low-probability region of the posterior predictive distribution of BBH tilt angles, with peaks that favour anti-aligned spins (though the distributions are broad). The tilt angles of GW200105 and GW191219\_163120 are unconstrained. Since the measurements are dominated by the properties of the primary in all binaries, the tilts measured at the point of merger are largely set following the first SN (assuming isolated binary evolution and that the BH formed first). Hence, the discrepancy in the populations may reflect differences in the pre-SN evolution or the distribution of natal kick velocities.

The peaks of the posterior tilt distributions in these events are in tension with canonical assumptions for BH natal kicks where velocities are either attenuated by post-SN fallback of material or modified in proportion with the BH mass \citep[cf.][]{Rodriguez16}. If real, this may be resolved via BHs that are imbued with kick velocities of comparable magnitudes to NSs \citep[see also][]{OShaugnessy17,Repetto17,Chruslinska18,Atri19,Fragione21}, extreme NS kick velocities \citep{Zhu21}, and/or extra tilt induced by stable mass transfer \citep{Stegmann21}. However, the broad distributions also extend to more moderate tilts within their 90 per cent confidence intervals. We also show that the distribution of kicks are likely isotropic, since preferentially polar kicks \citep[motivated by the observed correlation between the kick direction and spin axis of newly formed pulsars;][]{Johnston05,Yao21} are inefficient at producing tilt.

Independent of our considerations for the tilt angles, the early sample of NS-BH events and candidates display lower total binary masses and mass ratios than expected under typical assumptions for stellar evolution. These properties are consistent with an evolutionary pathway that features efficient stable mass transfer prior to the first SN, e.g. Model C in \citet{Broekgaarden21}. The rates predicted by this model also match the event-based rate of NS-BH mergers in \citet{Abbott21}. If this is a general property of binary evolution then the `classic' channel may be much less common than is typically assumed, with more massive stellar binaries merging before they become compact objects. We predict that future NS-BH detections in GW will continue to be distributed around a total binary mass of $\sim 7$\,M$_{\odot}$ and a mass ratio of $q \approx 3$ if this is the correct interpretation.

If the suggestion of highly tilted BHs in NS-BH mergers is real then the prospects for EM accompaniments are poor, since the amount of material that is stripped from the NS before merger is a function of the orbit-aligned spin component of the BH. Anti-aligned spins like the ones found at the peaks of the posterior distributions for GW200115, GW190917\_114630 and GW190426\_152155 are particularly detrimental to the prospects for multi-messenger observations of NS-BH binaries \citep[e.g.][]{Foucart18}. However, the weak constraints on key EM-predictive properties like $\chi_{\rm BH}$ and $q$ demonstrate the need for future studies that link GW signals to their expected EM counterparts. In tandem, these can provide much stricter constraints on the parameter space, helping to break degeneracies in the GW measurements.

We suggest that the evidence from the early LVK observations of NS-BH spins, masses and rates points towards a high mass transfer efficiency and strong isotropic BH natal kicks, which are both key ingredients for future population synthesis models.

\section*{Acknowledgements}

We thank the anonymous referee for careful consideration of this manuscript and for suggestions that improved the detail and scope of our study. We thank Ilya Mandel, Davide Gerosa, Floor Broekgaarden and Edo Berger for useful discussions on binary evolution physics, Chris Moore for help with the statistical interpretation, and Carl Rodriguez for providing simulation data. MN is supported by a Royal Astronomical Society Research Fellowship. MN and BG acknowledge funding by the European Research Council (ERC) under the European Union’s Horizon 2020 research and innovation programme (grant agreement No.~948381). AV acknowledges
the support of the Royal Society and Wolfson Foundation. This research has made use of data, software and/or web tools obtained from the Gravitational Wave Open Science Center (https://www.gw-openscience.org/ ), a service of LIGO Laboratory, the LIGO Scientific Collaboration and the Virgo Collaboration. LIGO Laboratory and Advanced LIGO are funded by the United States National Science Foundation (NSF) as well as the Science and Technology Facilities Council (STFC) of the United Kingdom, the Max-Planck-Society (MPS), and the State of Niedersachsen/Germany for support of the construction of Advanced LIGO and construction and operation of the GEO600 detector. Additional support for Advanced LIGO was provided by the Australian Research Council. Virgo is funded, through the European Gravitational Observatory (EGO), by the French Centre National de Recherche Scientifique (CNRS), the Italian Istituto Nazionale di Fisica Nucleare (INFN) and the Dutch Nikhef, with contributions by institutions from Belgium, Germany, Greece, Hungary, Ireland, Japan, Monaco, Poland, Portugal, Spain.

\section*{Data Availability}

The data used in this publication are available from the authors upon reasonable request.

The posterior samples for GW190426\_152155 are from GWTC-2 \citep{Abbott21b} and can be accessed at \href{https://dcc.ligo.org/LIGO-P2000223/public/}{https://dcc.ligo.org/LIGO-P2000223/public/} under sample set {\sc PrecessingSpinIMRHM}.

The posterior samples for GW190917\_114630 are from GWTC-2.1 \citep{GWTC-2.1} and can be accessed at \href{https://zenodo.org/record/5117703\#.YZYe6FOnzuR}{https://zenodo.org/record/5117703\#.YZYe6FOnzuR} under sample set {\sc PrecessingSpinIMRHM}.

The posterior samples for GW191219\_163120, GW200105 and GW200115 are from GWTC-3 \citep{GWTC-3} and can be accessed at \href{https://zenodo.org/record/5546663\#.YZYO41Ony8U}{https://zenodo.org/record/5546663\#.YZYO41Ony8U} under sample set {\sc C01:Mixed} for GW191219\_163120 and GW200105, and sample set {\sc C01:IMRPhenomXOHM:HighSpin} for GW200115 (for which no mixed samples were available).


\bibliographystyle{mnras}
\bibliography{main}




\bsp	
\label{lastpage}
\end{document}